\documentclass[10pt,letterpaper]{article}
\usepackage[top=0.85in,left=2.75in,footskip=0.75in]{geometry}

\usepackage{amsmath,amssymb}

\usepackage{changepage}
\usepackage{tabulary}
\usepackage[utf8x]{inputenc}

\usepackage{textcomp,marvosym}

\usepackage{cite}

\usepackage{nameref,hyperref}

\usepackage[right]{lineno}
\usepackage{tabularx}
\usepackage{microtype}
\DisableLigatures[f]{encoding = *, family = * }

\usepackage[table]{xcolor}

\usepackage{array}

\usepackage{tikz}
\usetikzlibrary{shapes, arrows}

\newcolumntype{+}{!{\vrule width 2pt}}

\newlength\savedwidth

\raggedright
\usepackage[aboveskip=1pt,labelfont=bf,labelsep=period,justification=raggedright,singlelinecheck=off]{caption}

\makeatletter
\renewcommand{\@biblabel}[1]{\quad#1.}
\makeatother

\usepackage{lastpage,fancyhdr,graphicx}
\usepackage{epstopdf}
\fancyheadoffset[L]{2.25in}
\fancyfootoffset[L]{2.25in}
\begin{document}
\vspace*{0.2in}

\begin{flushleft}
{\Large
\textbf\newline{Phase transitions and control measures for network epidemics caused by infections with presymptomatic, asymptomatic, and symptomatic stages} 
}
\newline
\\
Benjamin Braun\textsuperscript{*1\Yinyang},
Ba\c{s}ak Tarakta\c{s}\textsuperscript{2\Yinyang},
Brian Beckage\textsuperscript{3,4,5\Yinyang},
Jane Molofsky\textsuperscript{3\Yinyang}
\\

\bigskip
\textbf{1} Department of Mathematics,
University of Kentucky, Lexington, Kentucky, USA
\\
\textbf{2} Department of Political Science, Bo\u{g}azi\c{c}i University, Istanbul, Turkey
\\
\textbf{3} Department of Plant Biology, University of Vermont, Burlington, Vermont, USA
\\
\textbf{4} Department of Computer Science, University of Vermont, Burlington, Vermont, USA
\\
\textbf{5} Gund Institute for Environment, University of Vermont, Burlington, Vermont, USA
\\
\bigskip

%
%
\Yinyang These authors contributed equally to this work.

* benjamin.braun@uky.edu

\end{flushleft}
\section*{Abstract}
We investigate phase transitions associated with three control methods for epidemics on small world networks.
Motivated by the behavior of SARS-CoV-2, we construct a theoretical SIR model of a virus that exhibits presymptomatic, asymptomatic, and symptomatic stages in two possible pathways.
Using agent-based simulations on small world networks, we observe phase transitions for epidemic spread related to: 1) Global social distancing with a fixed probability of adherence. 2) Individually initiated social isolation when a threshold number of contacts are infected. 3) Viral shedding rate.
The primary driver of total number of infections is the viral shedding rate, with probability of social distancing being the next critical factor. 
Individually initiated social isolation was effective when initiated in response to a single infected contact. 
For each of these control measures, the total number of infections exhibits a sharp phase transition as the strength of the measure is varied.

\section*{Introduction}
The SARS-CoV-2 virus that has spread throughout the globe has created societal disruption and had a massive impact on global health \cite{forecastingCOVID}. 
With no known treatment, public policy and human behavior are currently the only tools that are available to mitigate the spread \cite{longevity}. 
A fundamental characteristic of SARS-CoV-2 is that after an individual is exposed, that individual passes through an extended presymptomatic stage followed by either an asymptomatic or symptomatic stage~\cite{covidreport}.
Our goal in this work is to construct a theoretical network disease model with these qualities and investigate phase transitions associated with three types of control measures.
While many models related to SARS-CoV-2 are designed to be forecasting tools, our study is intended as a contribution to the theoretical literature regarding qualitative aspects of control measures for viruses with these pathways of disease progression.

In the specific case of SARS-CoV-2, one control measure that has been used is government-mandated social distancing.
Different countries, and different states within the US, have implemented different approaches to this~\cite{covidreport, longevity}.
While most government plans include some social distancing, questions have arisen as to the efficacy of social distancing, how long social distancing should last and to what extent it is needed \cite{Hellewell:2020aa}.
A second control method involves individually-determined changes to social behavior, which work in concert with mandated social distancing to mitigate viral transmission~\cite{Bavel:2020aa, wise_zbozinek_michelini_hagan_mobbs_2020}. 
Individuals who live with an infected individual are being asked or required to quarantine for 14 days prior to interacting in the larger society~\cite{quar}.
One question is whether these individual responses of behavioral modification are sufficient to moderate epidemic spread and whether there are additive or non-additive effects when implemented with top-down government policy on social distancing~\cite{Bavel:2020aa}.  
A third type of control measure involves use of personal protective equipment to reduce the rate of viral transmission.
For example, mask usage has been found to be effective in this regard for SARS-CoV-2~\cite{lancetfacemasks,Syed855, BehavChanInfec}.

These real-world aspects of SARS-CoV-2 highlight the need for a more thorough understanding of the general behavior of viruses exhibiting multiple progressions of disease development.
With this as motivation, we develop a theoretical model in which we investigate how three types of control measures are associated with sharp phase transitions for the total number of infected individuals.
While modeling contacts can be done in mean-field, statistical, and metapopulation SIR models~\cite{FANELLI2020109761, gaeta2020simple,forecastingCOVID, geotemp, machine}, we use an agent-based model (ABM) on Watts-Strogatz small world networks~\cite{Watts:1998aa,Amaral11149, wattsStrog}.
Small world networks have connectedness properties that are found in real-world social networks and have been previously considered in epidemiological contexts~\cite{SIRntwmixing, epstein}.

The first control measure in our ABM is social distancing imposed on the network at a global scale.
Our model encodes this global social distancing as complete isolation of an agent from other agents.
The likelihood of social distancing is applied uniformly to all agents. 
The second control measure arises when agents have social connections that are infected and symptomatic~\cite{longevity, BehavChanInfec}. 
In this case, agents temporarily isolate from their contacts in the network if they are in contact with a sufficient number of symptomatic agents.
The third type of control measure is to alter the rate of viral spread, which reflects behavior such as use of personal protective equipment, e.g., masks~\cite{Syed855, BehavChanInfec}. 
We examine how each of these measures alone and in concert with each other influence the viral outbreak.

For each of these control measures, we ask the following questions:
\begin{enumerate}
    \item How does varying the strength of the control measure impact the total number of infections in an epidemic?
    \item If a control measure impacts the total number of infections, is there a phase transition associated with changes in strength of that control measure?
    \item How do these three control measures interact regarding their impact on total number of infections?
\end{enumerate}

\section*{Model and Parameters}

\subsection*{Agent-Based Model}
\begin{figure}[!h]
\centering
\caption{{\bf Agent Infection States.} Flow chart of infection pathways in the agent based model.}
\label{fig:model}
\includegraphics{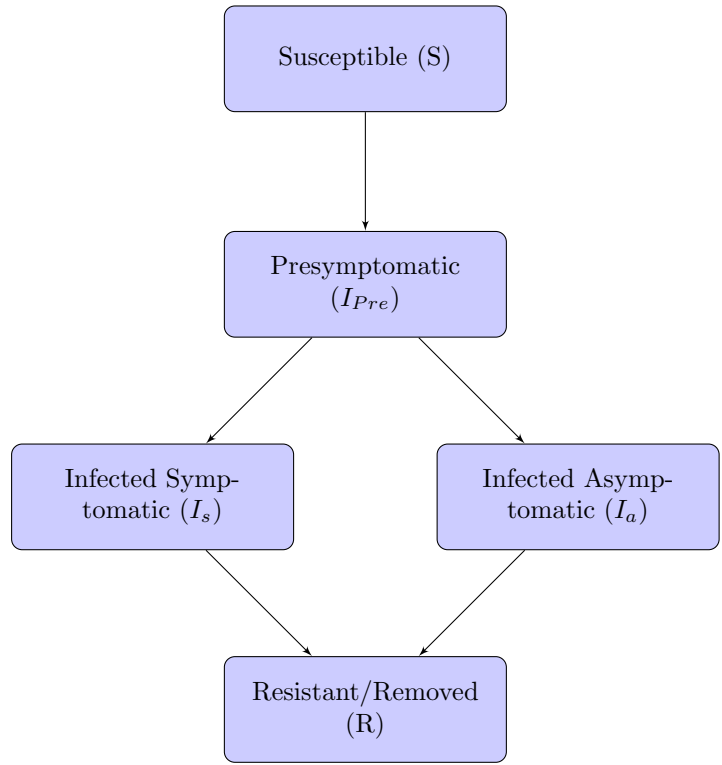}
\end{figure}

We develop an SIR, network-based,  agent-based model where agents pass through various infection states (Fig.~\ref{fig:model}). 
Agents pass through a presymptomatic infection state followed by either an asymptomatic infected stage or a symptomatic infected stage.
In our model, each agent carries an individual pathogen level that changes in response to contact with infected agents.
Initially, this level is set to 0 pathogen units for susceptible agents. At each time step (conceived as a day), if a susceptible agent has no infected contacts then their pathogen level does not change.
For each day that a susceptible agent has one or more infected neighbors, their pathogen level increases by a fixed fraction of the pathogen levels of their infected neighbors.
There is a global infection threshold that applies to all agents, which we fix at 25 pathogen units; in other words, the day after the pathogen level for an individual agent exceeds 25 units, that agent enters the presymptomatic infected state.
Once an agent enters the infected state, their pathogen level stays constant until they have reached the resistant/removed state, at which point it is reset to 0 units.
Model runs are initiated with a small number of infected agents, whose pathogen levels are initially set at 35 pathogen units, and the remainder of the agents are initially deemed susceptible. 

These initial and threshold values for the pathogen levels in our model are not based on real-world data, but rather were selected for simplicity to investigate general behavior of phase transitions under a mechanism of viral shedding with individualized pathogen levels.
Because our model does not use a transmission probability for each contact, but rather a viral shedding rate where each individual agent has varying levels of pathogen load, this model is well-suited to ABM simulations and less amenable to ODE-based deterministic analysis.

Once the individual pathogen level for an agent exceeds 25 units, that agent enters a presymptomatic infection stage, followed by either a symptomatic or asymptomatic stage (Fig.~\ref{fig:model}). 
The length of the presymptomatic stage is the same for all agents, and can be set to last one or more days. 
The lengths of the two possible main infection stages are set independently from each other, but are the same for all agents.
Following the main infection stage, the agent is either resistant or removed.

In addition to the infection stages, each agent is in one of two daily behavior states: socially distanced or not socially distanced.
The behavior state is reset each day.
If an agent is not socially distanced on a given day, then that agent can interact with any neighboring agent.
If an agent is socially distanced, the agent does not interact with any neighboring agents; in our theoretical model, social distancing is equivalent to self-quarantine.
Agents socially distance in a given day for one of two reasons.
A global social distance probability is set, which determines the chance that an agent will socially distance on a given day.
A local social distance threshold is set, and this value dictates individual responses to infected symptomatic neighbors.
If the number of infected symptomatic neighbors of an agent equals or exceeds this threshold, the agent will social distance independently of the global parameter.
This local threshold is the same for all agents.

\subsection*{Model Parameters and Simulations}

We implement our agent-based model on Watts-Strogatz (WS) networks. 
The WS networks can simultaneously demonstrate both high clustering and short average path length, and thus serve as effective approximations of social networks that are neither completely random nor regular~\cite{Watts:1998aa}. 
High clustering and short average path length allow for local interactions and more distant interactions to be incorporated~\cite{wattsStrog, Amaral11149, Chen_2007}, which are properties often found in real-world networks.
The WS small world network in our model is characterized by three parameters: number of nodes $N$, average node degree $K$ and rewiring probability. 
The rewiring parameter is used to determine the likelihood of rewiring each edge starting from a regular ring lattice.
A rewiring parameter of $0$ preserves the original ring lattice; a rewiring parameter of $1$ simulates a random network. 
We fix the number of nodes $N=500$ and the average degree $K=20$, which allows $\ln(N) \ll K \ll N$.
We then vary the rewiring probability among the values $\{0.05, 0.10, 0.25, 0.50\}$. 
For each of our four rewiring probabilities we construct $10$ networks on which to run simulations.

We define our three model parameters as follows:
\begin{enumerate}
    \item Social distance probability: the probability that an agent is socially distancing on any given day.
    \item Social distance threshold: the minimum number of infected symptomatic contacts required to cause an agent to social distance for that day.
    \item Viral shedding: the fraction of individual pathogen level that an infected agent passes to each of its contacts.
\end{enumerate}

We ran two sets of simulations over different parameter spaces.
Our primary simulation ran through ten networks for each set of parameters given in Table~\ref{table:paremeters}.
Based on the results of this primary simulation, we ran a secondary set of simulations over the refined parameter space given in Table~\ref{table:secondaryparameters} to provide a more detailed analysis of the phase transition behavior observed in the primary simulations.
The parameters for the secondary simulation were selected based on our analysis of the primary data using regression trees to identify critical variables and on the observed ranges where phase transitions were observed.

\begin{table}
\caption{
{\bf Primary Simulation Parameters}}
\begin{tabularx}{\textwidth}{|X|X|X|}
\hline
\textbf{Parameter}                   & \textbf{Description}                          & \textbf{Value} \\ \hline \hline
\textit{\textbf{Infection}}        &            &       \\\hline
Presymptomatic stage & incubation period after exposure         &   1, 3, 5    \\\hline
Pathogen infection threshold    & amount of virus to become infected    & 25 \\\hline

Asymptomatic infection stage            & period of infection without symptoms   &  5, 8, 10      \\\hline
Symptomatic infection stage            &  period of infection with symptoms  &  5, 8, 10      \\\hline
Chance symptomatic & 
 the probability an agent becomes symptomatic   & 0.25, 0.50, 0.75      \\
\hline \hline
\textit{\textbf{Network}}   &         &       \\\hline
Number of nodes (N)         & number of agents     &    500   \\\hline
Average node degree (K)            & the average number of contact for agents  &   20    \\\hline
Rewiring probability        & probability each edge is rewired  &   0.05, 0.10, 0.25, 0.50   \\\hline
Initial outbreak size & number of infected agents at the start of the run & 5 \\
\hline \hline
\textit{\textbf{Social Distance and Controls}}    &     &       \\\hline
Social distance probability      & the probability (out of 100) that an agent socially distances on a given day (reported as a percentage)   &  0, 10, 20, 30, 40, 50, 60, 70, 80, 90     \\\hline
Social distance threshold   & number of infected neighbors needed for an agent to social distance &   1, 2, 4, 8     \\\hline 
Viral shedding &  the amount of pathogen level that the infected agents shed &
  1\%, 5\%, 25\% \\
\hline

\end{tabularx}
\label{table:paremeters}
\end{table}

\begin{table}
\caption{
{\bf Secondary Simulation Parameters}}
\begin{tabularx}{\textwidth}{|X|X|}
\hline
\textbf{Parameter}                           & \textbf{Value} \\ \hline \hline

\textit{\textbf{Infection}}        &             \\\hline
Presymptomatic stage &   3   \\\hline
Pathogen infection threshold & 25 \\\hline
Asymptomatic infection stage             &   8      \\\hline
Symptomatic infection stage             &  8    \\\hline
Chance symptomatic  & 0.25, 0.50, 0.75       \\    
\hline \hline
\textit{\textbf{Network}}          &       \\\hline
Number of nodes (N)              &    500   \\\hline
Average node degree (K)              &   20    \\\hline
Rewiring probability        &   0.10   \\\hline
Initial outbreak size  & 5 \\
\hline \hline
\textit{\textbf{Social Distance and Controls}}    &       \\\hline
Social distance probability (as percentage)       &  60, 61, 62,\ldots, 78, 79, 80     \\\hline
Social distance threshold   &  1, 2, 3, 4, 5    \\\hline 
Viral shedding &
  5\%, 10\%, 15\%, 20\% \\
\hline

\end{tabularx}
\label{table:secondaryparameters}
\end{table}

\section*{Results}

\subsection*{Regression Tree}

\begin{figure}[!h]
\caption{{\bf Regression Tree for Total Number of Infections.} This tree identifies the input features with strongest influence on total number of infections. 
Each box contains the percentage of observations and associated mean viral load of agents.
For example, in the upper-left box, 67\% of the simulations had viral shedding below 15\%, and for those simulations the mean number of infections was 46.}
\label{fig:regtree}
\includegraphics[width=\linewidth]{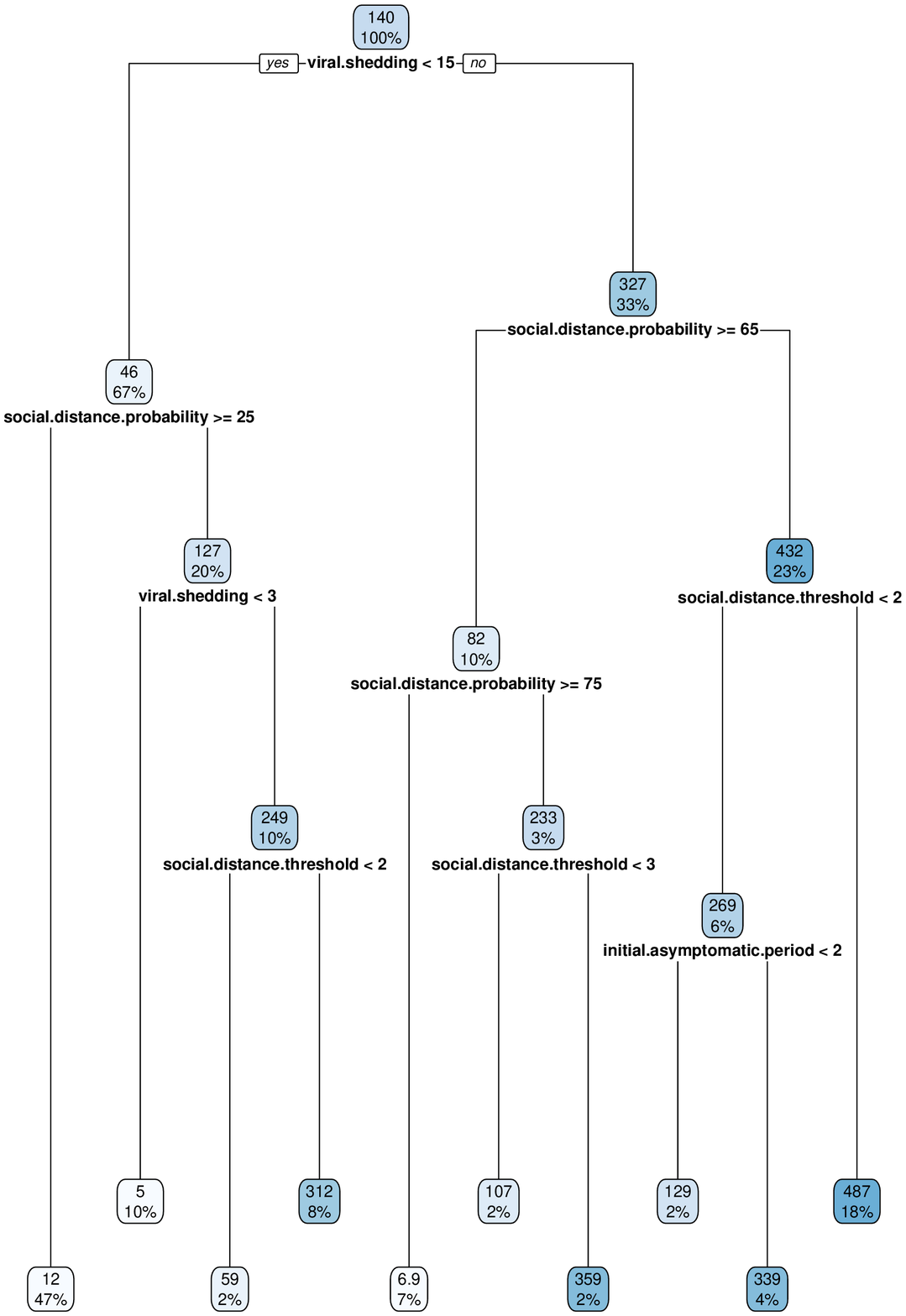}
\end{figure}

We used a regression tree to partition the variation in final number of infected nodes across model parameters and runs in our primary simulation~\cite{regTrees}.
Reductions in viral shedding were associated with the primary partition in the regression tree in Fig.~\ref{fig:regtree}. 
Viral shedding below 15\% compared to a value of 25\% were associated with a mean number of infections of 46 out of 500 agents.
Reduced viral shedding with  social distancing probability over 25\% led to  overall infection of approximately 2\% of the agents. 
If the overall viral shedding is reduced dramatically to 5\%, even without additional social distancing of any type, less than 1\% of the population becomes infected.

Achieving low levels of infections in populations without reducing viral shedding requires significantly higher levels of global social distancing, where each agent has at least a 75\% chance of social distancing each day; this results in an approximately 1\% infection rate among agents.  
If each agent has less than a 75\% chance of social distancing each day, the total infection rates for the populations are much higher; these range from a low of 21\% (if individuals self-isolate in response to one infected social contact) all the way up to 97\% with low levels of any type of social distancing.

Thus, with a higher level of viral shedding, it becomes important to have agents self-isolate when a contact becomes symptomatic.
Even if this occurs, the infection rate in the population is an order of magnitude higher (10\% vs. 1\%) than if the viral shedding is reduced. 
Failure to achieve this strict social distancing in response to an infected social contact results in a wide-spread outbreak with approximately 62\% of the agents infected.

\subsection*{Phase transitions}

\begin{figure}[!h]
\caption{{\bf Viral Shedding.}} 
Viral shedding vs. total number of infected agents for different levels of social distance thresholds (1 through 5) and for social distance probabilities varying from 60\% to 80\% at 1\% intervals.
A clear phase transition exists at viral shedding 10-15\% across the levels of social distance thresholds and social distance probabilities.
\label{fig:shedding}
\includegraphics[angle=-90,width=\linewidth]{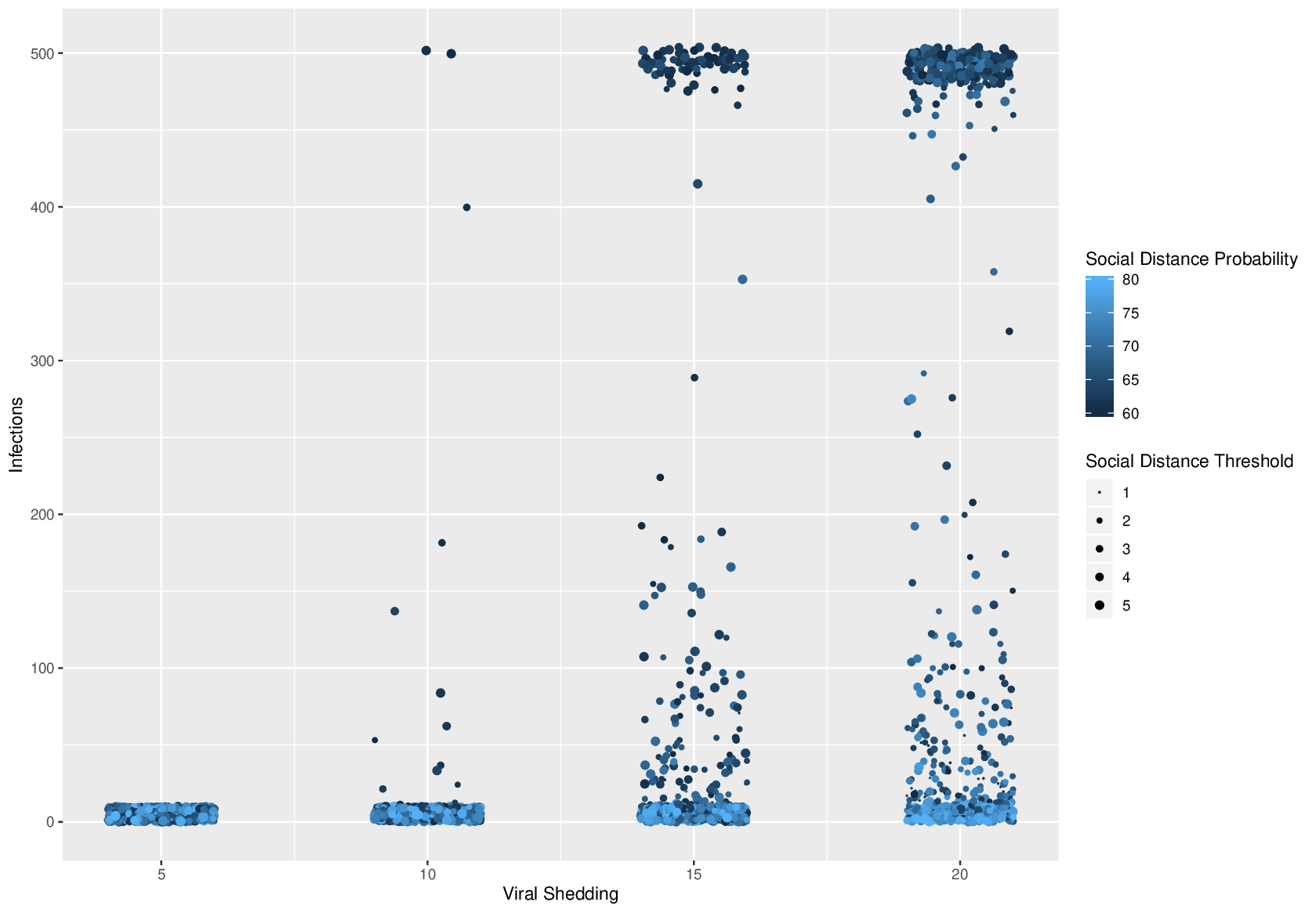}
\end{figure}

\begin{figure}[!h]
\caption{{\bf Social Distance Threshold.}
Social distance threshold vs. total number of infected agents for different levels of viral shedding (5\% to 20\%) and for social distance probabilities varying from 60\% to 80\% at 1\% intervals. 
A clear phase transition exists at a social distance threshold of 1, across all levels of social distance thresholds and social distance probabilities.
}
\label{fig:socdistthreshold}
\includegraphics[angle=-90,width=\linewidth]{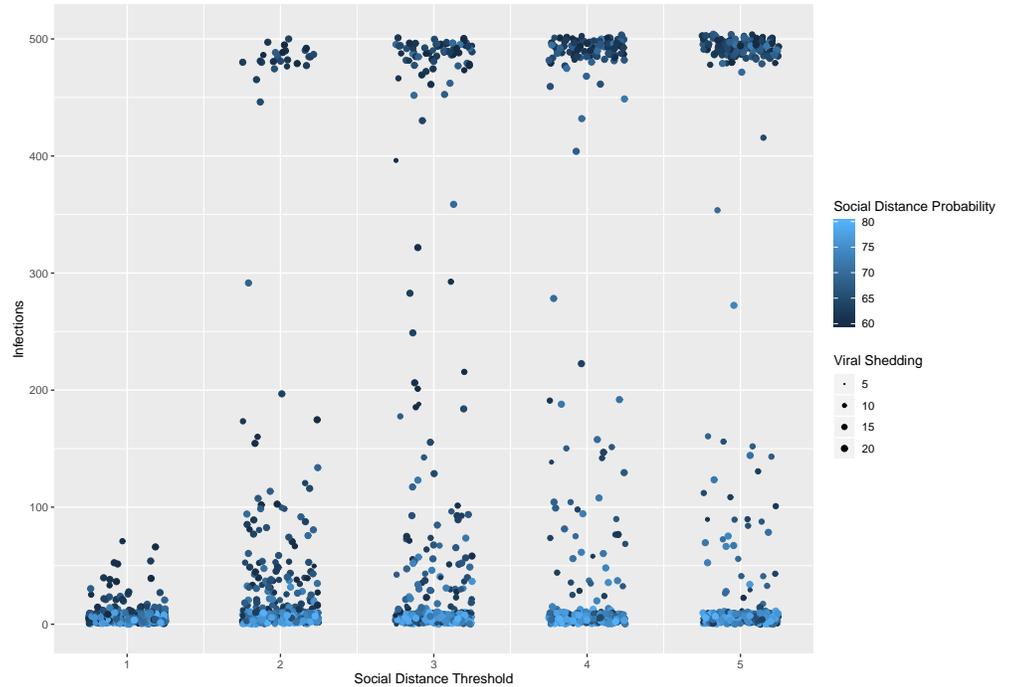}
\end{figure}

\begin{figure}[!h]
\caption{{\bf Social Distance Probability.}
Social distance probability vs. total number of infected agents for different levels of viral shedding (5\% to 20\%) and for social distance threshold varying from 1 to 5.
A phase transition exists around a social distance probability of 73-74\%, across the levels of social distance threshold and viral shedding.
}
\label{fig:socdistchance}
\includegraphics[angle=-90,width=\linewidth]{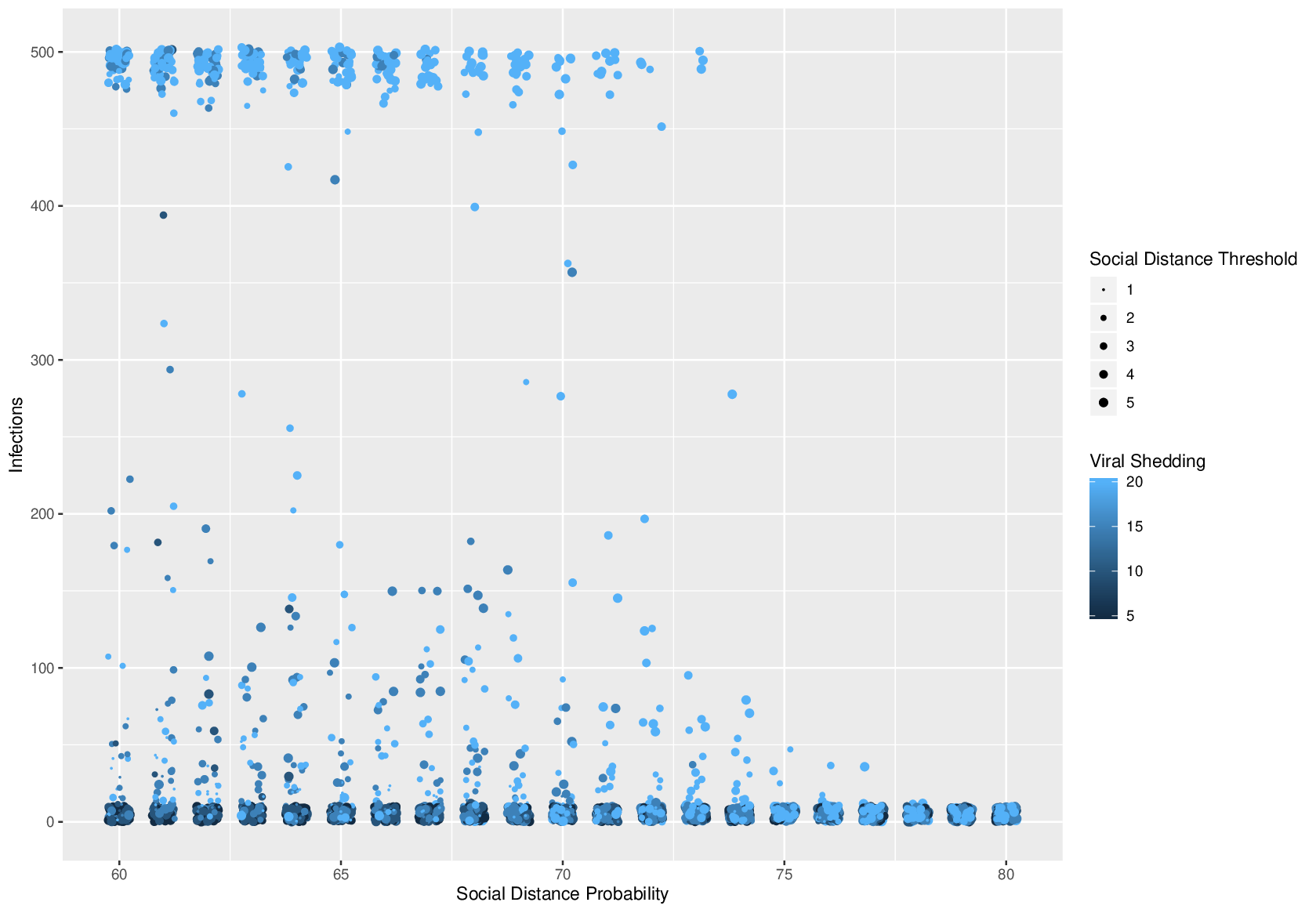}
\end{figure}

Because our goal is to understand the behavior of phase transitions regarding total number of infections in our model, we conducted secondary simulations on a refined parameter space based on the results of our regression tree analysis. 
In these simulations, we observed sharp phase transitions in the total number of infections as a function of all three control methods.
These transitions are shown in Fig.~\ref{fig:shedding}, Fig.~\ref{fig:socdistthreshold}, and Fig.~\ref{fig:socdistchance}.
In these figures, the maximum number of possible infections is 500, as there are 500 nodes in our networks. 

In Fig.~\ref{fig:shedding}, a phase transition exists between viral shedding of 5\% and 10\%, across all levels of social distance thresholds and social distance probabilities.
In Fig.~\ref{fig:socdistthreshold}, a phase transition exists at a social distance threshold of 1, across all levels of social distance probabilities and viral shedding.
If the social distance threshold parameter is 2 or more, then it is possible to have epidemics that infect the entire population.
In Fig.~\ref{fig:socdistchance}, a phase transition exists around a social distance probability of 73-74\%, across all levels of social distance threshold and viral shedding.
If the social distance probability is 74\% or more, then our simulations end with a small number of infected agents.

\subsection*{Number of Infections and Length of Epidemic}

\begin{figure}[!h]
\caption{{\bf Total infections across social distance probability and viral shedding.}
Social distance probability vs. viral shedding, with social distance threshold and total number of infections indicated by color and size.
Larger numbers of infections occur with low social distance probability and high viral shedding rates.
}
\label{fig:thresholdshed}
\includegraphics[angle=-90,width=\linewidth]{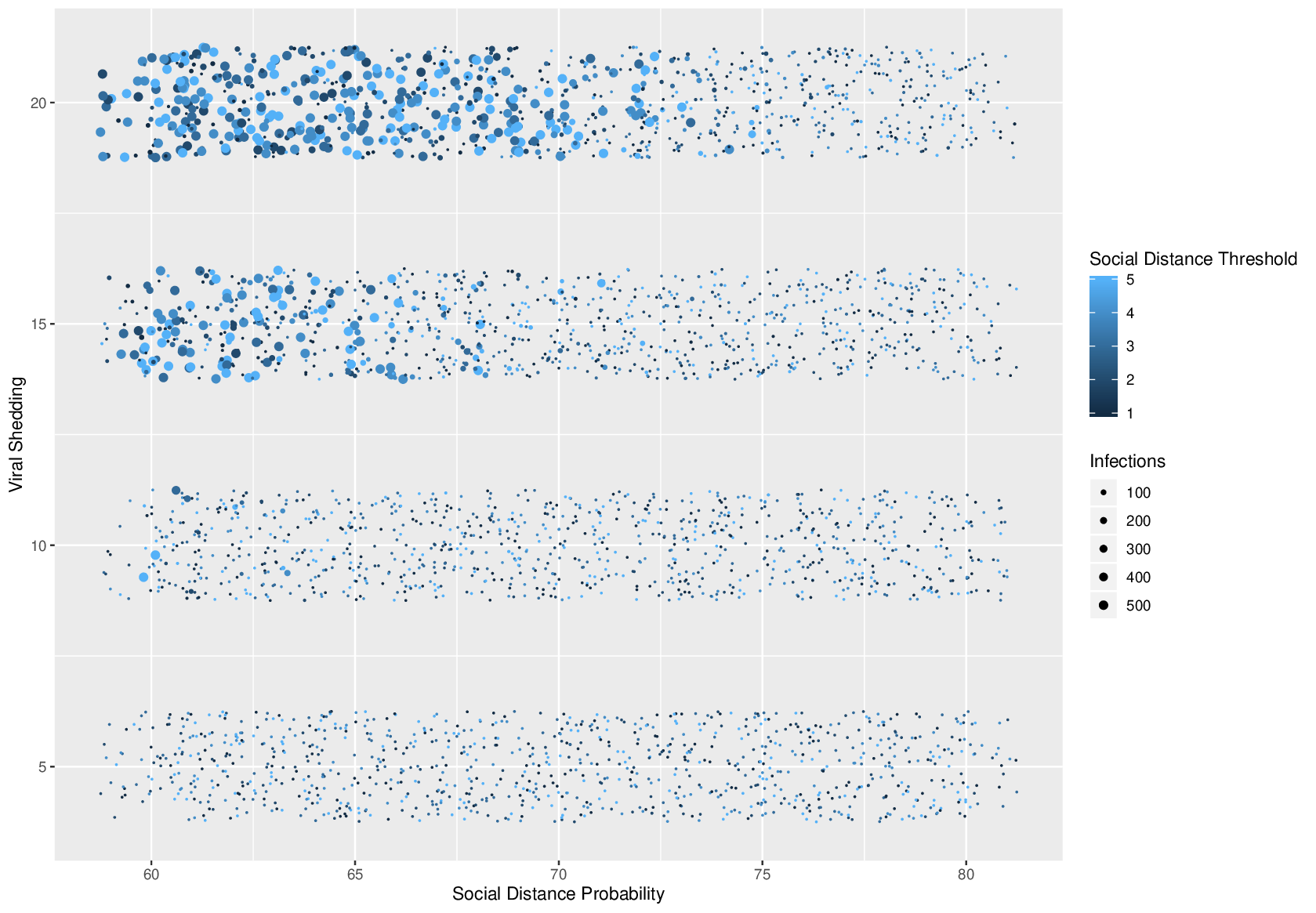}
\end{figure}

Given the regression tree analysis of our primary simulations, it is clear that viral shedding and social distance probability play key roles.
In our secondary simulations over a refined parameter space, this becomes more clear.
In Fig.~\ref{fig:thresholdshed}, we observe additional confirmation of the regression tree findings that the main driver of total number of infections is the viral shedding rate, with social distance probability being the next critical factor.
Specifically, simulations with large total infections cluster to the upper left of the plot, where viral shedding rates are higher and social distancing is enacted by approximately 60\% of agents.

\begin{figure}[!h]
\caption{{\bf Social Distance Probability and Viral Shedding.}
Social distance probability vs. viral shedding, with social distance threshold and length of epidemic indicated by color and size.
Longer outbreaks occur with low social distance probability and high viral shedding rates.
}
\label{fig:chanceandshedding}
\includegraphics[angle=-90,width=\linewidth]{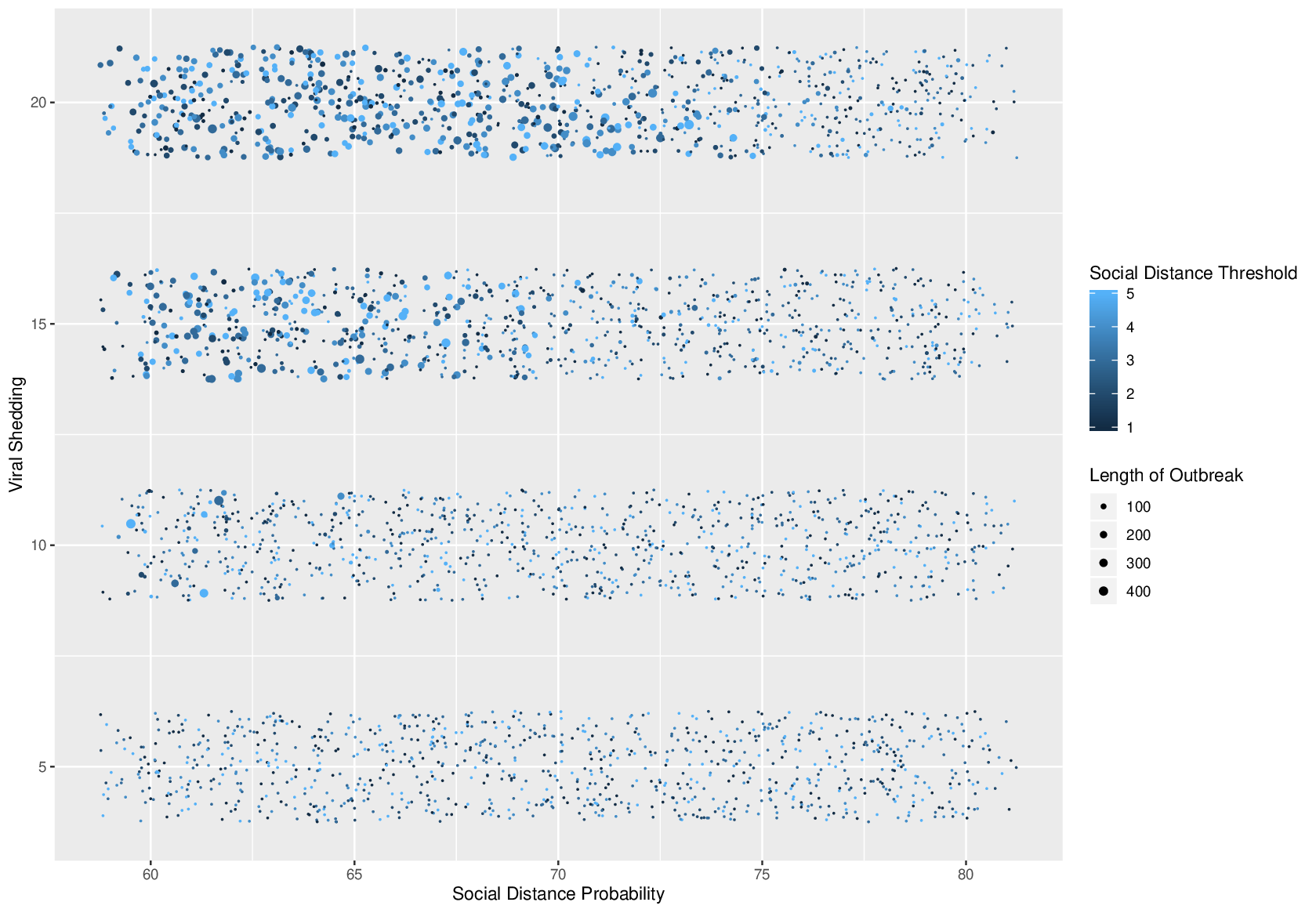}
\end{figure}

\begin{figure}[!h]
\caption{{\bf Total Infections and Length of Epidemic}
Length of epidemic vs. total number infected with social distance probability, social distance threshold , and viral shedding indicated by color, size, and symbol.
Data from secondary simulations.
}
\label{fig:infectionsandlength}
\includegraphics[angle=-90,width=\linewidth]{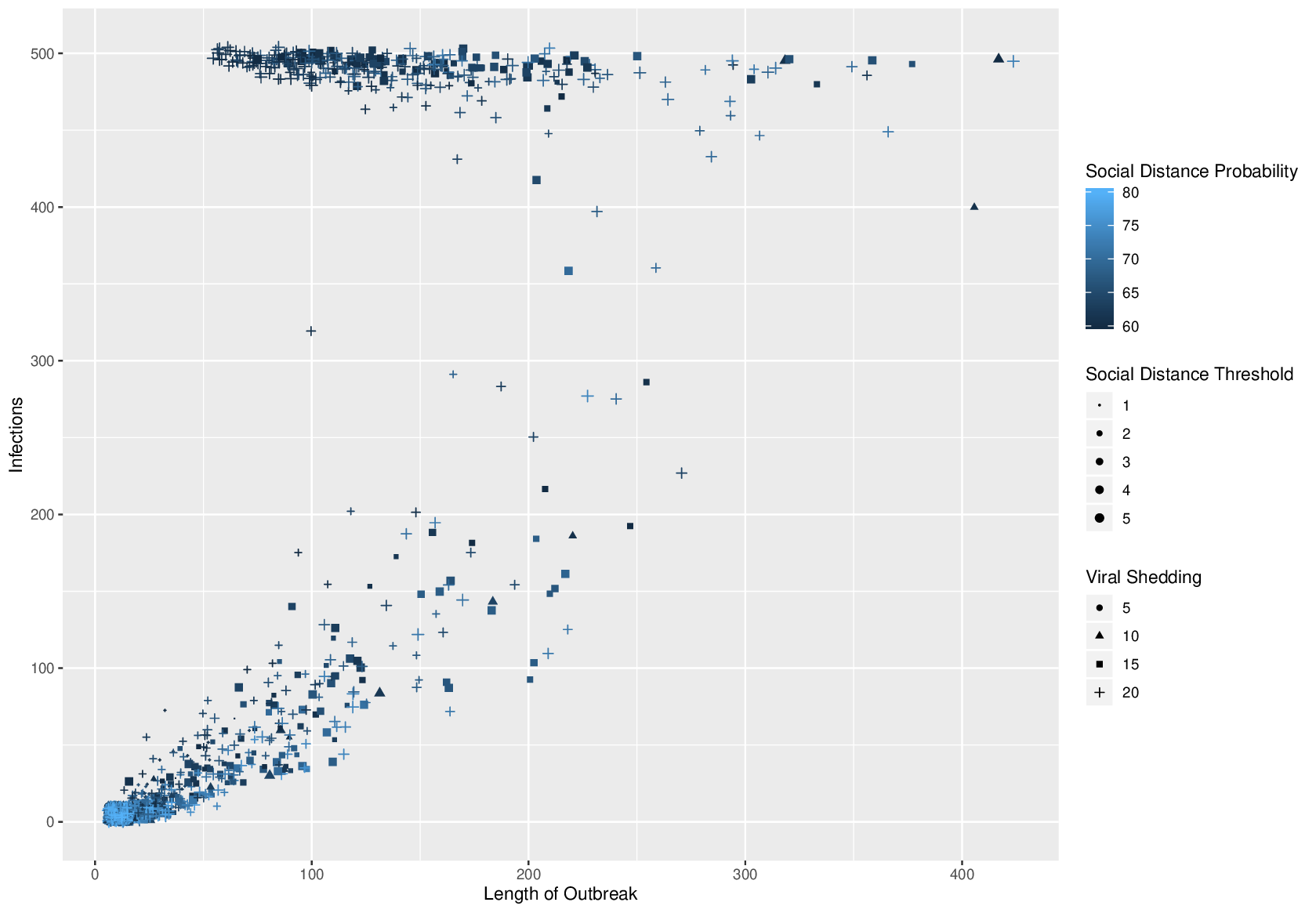}
\end{figure}

There is also a clear interaction between the social distance probability and viral shedding parameters and the resulting number of infected agents and the length of the epidemic. 
These interactions are shown in Fig.~\ref{fig:chanceandshedding} and Fig.~\ref{fig:infectionsandlength}.
In Fig.~\ref{fig:chanceandshedding}, there is clustering of long epidemics when the probability is near 60\% and the viral shedding rate is high.
As the social distance probability increases to 80\% and the viral shedding rate decreases, there is a phase transition where simulations result in outbreaks of short duration.
In Fig.~\ref{fig:infectionsandlength}, most infections result in either a limited outbreak (less than 125 out of 500 agents) or almost all agents infected.
As the social distance probability is increased from 60\% to 80\%, the length of the epidemics increase while remaining limited in total number of infections before sharply transitioning to a high number of infections during a return to short epidemic lengths.

\section*{Discussion}

Mathematical modeling can provide tools to better understand epidemic dynamics and can vary from purely theoretical to more data driven and predictive~\cite{epidemicsonnetworks}.   
While a simple model such as this one should not be used to make policy recommendations, it can provide a framework for empirical investigation and specific hypothesis testing related to social networks of smaller size exhibiting small world characteristics, such as those seen in college settings~\cite{college}.
Here we use our theoretical model to investigate how different control methods impact the total number of infections in an epidemic on a network caused by a virus with a presymptomatic stage and both asymptomatic and symptomatic pathways.
We specifically examine three main control measures that can be taken to reduce epidemic spread: 1) Global social distancing with a fixed probability of adherence. 
2) Individually initiated social isolation when a threshold number of contacts are infected.
3) Reduction of viral shedding.
We observe sharp phase transitions in the total number of infected agents as the strength of each of these control measures are varied.

To examine the full potential for global social distancing, we consider a wide range of possible scenarios varying from no social distancing to strong adherence to social distancing (90\% of agents). 
When considering the relationship between our theoretical model and real-world contexts, the two extreme scenarios are easy to envision (zero social distancing is business as usual and 90\% is all but non-essential businesses closed), while more moderate social distancing scenarios are harder to translate into direct societal actions.
Nevertheless, we observe in our small-world models a clear phase transition associated with global social distancing.
In general, a global social distancing probability below 65\% results in a wide-spread epidemic, while a global social distancing probability above 75\% limits the epidemic to a dramatically lower number of total infections.  We also found that social distance probabilities that approached the threshold from below resulted in prolonged epidemics while with low overall infection rates.
For our secondary simulations over a refined parameter space, in the absence of other control measures we observe that there is a phase transition for total infections that occurs as the percentage of agents socially distancing changes from 73\% to 74\%.

Individual behavior taken during a pandemic can greatly affect the dynamics of disease spread.
For example, for SARS-CoV-2, the most commonly recommended guideline after contact with an infected individual is 14 days of self-isolation to avoid exposing other individuals~\cite{Bavel:2020aa, quar, perceptions}.
However, despite these official guidelines, self-isolation following exposure requires that infected individuals inform their contacts and that exposed individuals voluntarily comply.
Thus, from a theoretical perspective it is important to understand how different self-isolation behaviors following contact with an infected agent impact epidemic spread.
In our model, we consider self-imposed social distancing as highly responsive to an agent's short-term perceptions regarding infection risks within their community.
Thus, self-imposed social distancing/isolation occurs only on the days when the agent has sufficiently many symptomatic contacts in the network. 
Interestingly, for self-isolation to significantly decrease the total number of infections in our model, an extreme level of responsiveness was needed by the agents involved; in our model, it was necessary for self-isolation to occur following exposure to only one infected agent. 
If self-isolation occurred only after contact with two or more symptomatic agents on the same day, the effect on disease spread was minimal.
Our findings also support the well-known fact that real-world contact tracing following an individual's positive test is critical for limiting the spread of the infection~\cite{Hellewell:2020aa}.
An important difference between our theoretical model and viruses such as SARS-CoV-2~\cite{anast, geotemp} is that, in the real world, individuals who have come in contact with an infected individual are not aware of their exposure.

Our theoretical reduction of viral shedding is motivated by behaviors such as mask wearing or other use of personal protective equipment~\cite{Syed855}.
While in real-world contexts individual responsiveness to recommended government actions are highly variable~\cite{perceptions}, a decrease in viral shedding rate can be achieved through use of protective equipment~\cite{lancetfacemasks}.
When the viral shedding in our model was set at a high shedding rate of 25\%, global social distancing was required to be greater than 80\% to control the outbreak, resulting in an approximately 1\% infection rate in the population.  
Other less stringent social distancing conditions result in a viral infection rate between 25\% and 97.5\%.
With a moderate rate of viral shedding, the social distance threshold at which someone decides to self-isolate after coming into contact with an infected individual becomes much more important. 
In our model, if the social distance threshold is set to 1 (agents self-isolate after coming into contact with at least one infected agent), then the final total infection rate in the population is approximately 12\%.
However, if the behaviorally induced social distancing does not take place or takes place at a higher threshold, then the total number of infections is much larger with the overall infection rate in the population approximately 62\%.
If the viral shedding rate is very low, then the epidemic does not spread and a low total number of infections is observed.
Thus, there is a sharp phase transition as the viral shedding rate moves from 5\% to 10\%.

An important observation regarding these phase transitions is the relatively extreme values at which they occur, e.g., a high social distancing probability, a low social distancing threshold, and a low viral shedding rate.
These values are very high and low both within the context of our model and of real-world epidemics that motivate our model.
It would be of interest to investigate whether or not, given an arbitrary set of values, a specific network and selection of parameters could be found for which phase transitions occur near these values.
Alternatively, if no such network and choice of parameters exist, it would be of interest if a more rigorous theoretical description could be given of the mechanism preventing this occurrence.

\section*{Conclusion}

We develop an agent-based model of epidemic spread on Watts-Strogatz small world networks, where infected agents pass through a presymptomatic stage followed by either an asymptomatic or symptomatic stage.
We consider the impact of three control measures on the total number of infected agents, with regard to both phase transitions and efficacy.
The three control methods we consider all generate sharp phase transitions in the total number of infections as the strength of the method varies.
Social distancing controls in this model exhibit a phase transition regarding total number of infections, either when imposed globally or when based on individual response to infected contacts.
Individually-enacted social distancing in the form of temporary self-isolation must be immediately enacted if a social contact is known to be infected in order to halt the spread of an epidemic.
Reductions in viral shedding lead to significant reductions in the size of the final infected population.

\section*{Supporting information}


\paragraph*{S1 File.}
\label{S1_Software}
{\bf NetLogo code and experimental data.} https://github.com/braunmath/social-distance-effects-covid19

\section*{Acknowledgments}
This project began during the workshop ``Understanding and Exploring Network Epidemiology in the Time of Coronavirus'', organized by the University of Maryland's COMBINE program in network biology and the University of Vermont's Complex Systems Center in April 2020. BB and JM acknowledge the Networks in Ecology workshop supported by the University of Vermont's Presidents office. 
The authors would like to thank the organizers Michelle Girvan, Daniel Serrano, Juniper Lovato, Anshuman Swain, and Nick Mennona.

\nolinenumbers

%
%
%


\end{document}